\documentclass[twocolumn,prb,showpacs]{revtex4}
\usepackage{amsmath}
\usepackage{amssymb}
\usepackage{subfigure}
\usepackage{graphicx}
\usepackage{dcolumn}
\usepackage{bm}
\usepackage{setspace}
\makeatletter

\begin{document}

\title{Specific heat of underdoped cuprates: RVB versus Fermi arcs}

\author{J. P. F. LeBlanc$^1$}
\email{leblanc@physics.uoguelph.ca}
\author{E. J. Nicol$^1$}%
\author{J. P. Carbotte$^2$}
\affiliation{$^1$Department of Physics, University of Guelph,
Guelph, Ontario N1G 2W1 Canada} 
\affiliation{$^2$Department of Physics and Astronomy, McMaster
University, Hamilton, Ontario L8S 4L8 Canada}
\date{\today}
\begin{abstract}
A recent microscopic model of the pseudogap state, based on the resonating valence bond (RVB) spin liquid, has provided a simple ansatz for the electronic self energy in which a gap forms on the antiferromagnetic Brillouin zone as the limit of a Mott insulator is approached in the underdoped regime.  Here, the ansatz is employed to calculate the electronic specific heat when a superconducting gap is also included.  We find qualitative agreement with all experimental observations in the underdoped regime of the cuprates.  We explore the relationship of the theory to two other purely phenomenological approaches, the nodal liquid and the Fermi arc model, and provide justification for their use on experimental data in light of this microscopic RVB theory.
\end{abstract}

\maketitle

Simplified models, such as BCS theory, which ignore much of the complicated details associated with actual metals, have had an enormous impact on our understanding of the superconducting state.  A generalization of BCS to include the $d$-wave symmetry of the superconducting gap in the cuprates has taken us a long way in understanding the overdoped and optimally doped part of their phase diagram.  However, it has failed so far to provide even a first qualitative picture of the observed properties as the Mott transition is approached in the underdoped regime.  Some additional essential element is still missing which has been widely recognized as associated with the opening of a pseudogap, although the exact nature of the phenomenon remains unknown.  Recently, Yang, Rice and Zhang \cite{yrz:2006} have provided a simple model for the self energy in the pseudogap phase from which the coherent part of the electronic Green's function can be constructed.  The work is based  on results for a resonating valence bond spin liquid and contains a gap which is formed on the antiferromagnetic Brillouin zone (AFBZ) as the doping is reduced and the transition to a correlation-induced Mott insulating state is approached from the metallic side.  Since its appearance in 2006, the model has had considerable success in understanding some aspects of Raman \cite{valenzuela:2007} and optical properties \cite{illes:2009} of the underdoped cuprates and has also been applied to angle-resolved photoemission (ARPES) data.\cite{yrz:2008}

In view of these developments, it is very important to test the YRZ model on other data.
In this paper, we consider the electronic specific heat, which has long been known to show anomalous properties \cite{loram:1994,loram:1998,loram:2001} not describable within simple BCS theory. Lacking the existence of a well-developed and accepted microscopic theoretical framework, Loram et al. \cite{loram:1998,loram:1994,loram:2001} included a depression of the electronic density of states (DOS) near the Fermi energy to analyze their early specific heat work.
More recently, a similar analysis applied to optical data \cite{hwang:2008,hwang:CM:2008} has also yielded new insights, including clarification of temperature-dependent Fermi arcs.
More sophisticated, but still purely phenomenological, approaches to the pseudogap phase have appeared, including the idea of a nodal liquid \cite{kanigel:2006} and of temperature-dependent Fermi arcs \cite{kanigel:2006,storey:2008,storey:CAP:2008,storey:2007,norman:2007}.  In these models, the pseudogap exhibits $d$-wave symmetry and forms on the Fermi surface.  At high temperatures only the antinodal region is gapped while the remaining ungapped arc length about the nodal direction is proportional to temperature.  The limit of a nodal liquid is when the pseudogap is taken to form over the entire Fermi surface. \cite{kanigel:2006,norman:2007}  In comparison to these models, the YRZ approach is more securely based in microscopic theory.  It is also profoundly different in that the pseudogap forms on the AFBZ.  
After presenting our results, we will provide an analysis of how YRZ relates to both the nodal liquid and Fermi arc model.  

There are of course many other theoretical approaches to the pseudogap phase, for example, preformed pairs \cite{emery:1995}, existing below an energy scale set by the onset temperature, $T^*$, with phase coherence taking hold only at the lower superconducting $T_c$, or as extensions of BCS theory to include the formation of finite momentum pairs which persist above $T_c$.~\cite{chen:2001}
Another class of theories involves competing orders, such as $d$-density waves \cite{chakravarty:2001}, which set in at $T^*$ and can coexist with superconducting order below $T_c$.

\begin{figure}
\includegraphics[width=0.95\linewidth]{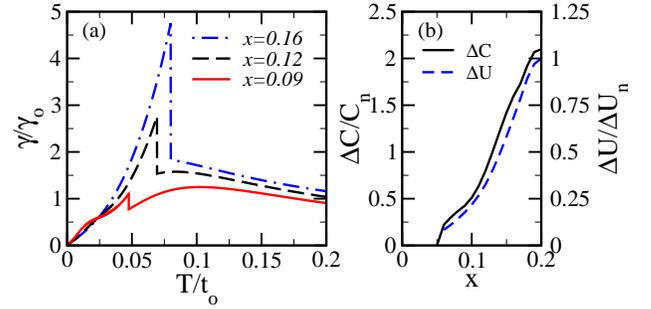}
\caption{\label{fig:fig1} (Color online) (a) Plot of $\gamma/\gamma_{0}$ vs
  $T/t_0$ for the YRZ model with $x=0.16$, 0.12 and 0.09.  (b) Plot of normalized jump, $\Delta C/ C_{\rm n}= \Delta C/\gamma (T_c)T_c$, 
  and normalized internal energy, $\Delta U/\Delta
  U_{\rm n}$ as functions of doping, $x$.}
\end{figure}

In the YRZ model, both the superconducting gap, $\Delta_{\rm sc}$, and
the pseudogap, $\Delta_{\rm pg}$, have a $d$-wave $k$-space dependence
described by: $\Delta_{\rm sc}=\frac{\Delta_{\rm sc}^{0}}{2}(\cos
k_xa -\cos k_ya)$ and  $\Delta_{\rm pg}=\frac{\Delta_{\rm
    pg}^{0}}{2}(\cos k_xa -\cos k_ya)$, with $a$ the lattice constant.
For a doping $x$, the YRZ model is described by a propagator,
\begin{eqnarray}
G(\boldsymbol{k},\omega,x)=
\sum_{\alpha=\pm}  {W^{\alpha}_{\boldsymbol{k}}}/[
{\omega-E^{\alpha}_{\boldsymbol{k}}-\Delta_{\rm sc}^{2}/(\omega+E^{\alpha}_{\boldsymbol{k}})}],
\label{eq:sc}
\end{eqnarray}
where $E_{\boldsymbol{k}}^ \pm  = \frac{{\xi_{\boldsymbol{k}}  - \xi_{\boldsymbol{k}}^0 }}{2} \pm E_{\boldsymbol{k}}$ , $
E_{\boldsymbol{k}} = \sqrt {\tilde{\xi_{\boldsymbol{k}}}^2  + \Delta
  _{\rm pg}^2 } $,
$\tilde{\xi_{\boldsymbol{k}}}  = (\xi_{\boldsymbol{k}}  + \xi_{\boldsymbol{k}}^0 )/2$
and $W_{\boldsymbol{k}}^ \pm   = \frac{g_{t}(x)}{2}\left( {1 \pm
  \frac{{\tilde{\xi_{\boldsymbol{k}}}  }}{E_{\boldsymbol{k}}}}
\right)$, where $g_{t}(x)$ weights the coherent part.~\cite{yrz:2008}  The energy dispersion $\xi_{\boldsymbol{k}} = - 2t(\cos
k_xa  + \cos k_ya ) - 4t^{\prime} \cos k_xa \cos k_ya - 2t''(\cos
2k_xa  + \cos 2k_ya )-\mu_p$ includes hopping out to 3$^{\rm rd}$
nearest neighbour, while $\xi_{\boldsymbol{k}}^0  =  - 2t(\cos k_xa  +
\cos k_ya)$ is the first nearest neighbour term, which determines the
placement of the pseudogap off the Fermi surface, coinciding with the
AFBZ boundary.  These energy dispersions contain doping dependent
coefficients: $t(x)=g_{t}(x)t_{0}+3g_{s}(x)J\chi/8$, $t^{\prime
}(x)=g_{t}(x)t_{0}^{\prime }$, and $t^{\prime\prime
}(x)=g_{t}(x)t_{0}^{\prime \prime }$,  where $g_{t}(x)=2x/\left(
1+x\right)$ and $g_{s}(x)=4/(1+x)^{2}$ are the Gutzwiller factors.
The dispersion here uses $\mu_p$ as an effective chemical potential or
Fermi level at $T=0$, determined by the Luttinger sum rule. Values of
other parameters in the dispersion were taken from Ref.~\cite{yrz:2006} to be: $t^{\prime}/t_{0}=-0.3$, $t^{\prime\prime}/t_{0}=0.2$,
$J/t_{0}=1/3$, and $\chi=0.338$; while the optimal superconducting gap
$\Delta_{\rm sc}^{0}$ was chosen to give an effective optimal $T_{c}$
around 90K for a ratio of $2\Delta^0_{\rm sc}(T=0)/k_{\rm
  B}T_c=6$,\cite{pasupathy:2008} for units of $t_{0}=0.1$ eV, where
$\Delta^0_{\rm sc}(T)$ is the gap amplitude given in BCS
theory at temperature, $T$.

From the YRZ propagator of Eq.~(\ref{eq:sc}), one can extract the YRZ spectral function and see that there are 
four energy branches, given by the energies, $\pm E_{\rm sc}^{\alpha}$,
where  $E^{\alpha}_{\rm sc}=\sqrt{(E_{\boldsymbol{k}}^{ \alpha })^2  +
  \Delta _{\rm sc}^2 }$.
One can straightforwardly calculate the specific heat, $C(T)$, from the entropy, $S$, given at temperature, $T$, by the standard formula summed over the four energy branches which reduces to 
\begin{eqnarray}
       S =  - 2{\rm{k}}_{\rm{B}} \sum\limits_{\alpha=\pm}\sum\limits_{\boldsymbol{k}} { {W^{ \alpha}_{\boldsymbol{k}}  \left\{ {f(E_{\rm sc}^{\alpha}  )\ln [f(E_{\rm sc}^{\alpha}  )] } \right.} } \nonumber\\
 {   \left. { + f( - E_{\rm sc}^{\alpha}  )\ln [f( - E_{\rm sc}^{\alpha}  )]} \right\} },
\label{eq:entropy}
\end{eqnarray}
where $f$ is the Fermi function and $k_{\rm B}$, the Boltzmann
constant.  The temperature dependence enters through  both $f$ and the
temperature-dependent superconducting gap.
\begin{figure}
\includegraphics[width=0.95\linewidth]{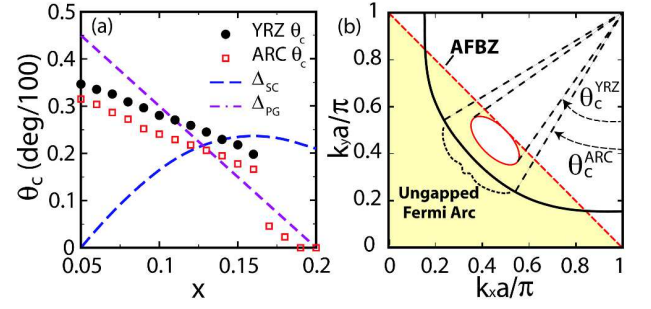}
\caption{\label{fig:thetas}(Color Online) (a) The phase diagram via
  the magnitudes of the gaps $\Delta^{0}_{\rm
    sc}(x)/t_0=0.24(1-82.6(x-0.16)^2)$ and $\Delta^{0}_{\rm
    pg}(x)/t_0=0.6(1-x/0.2)$. Also, the values of the critical angles,
  $\theta^{\rm ARC}_c$ and $\theta_c^{\rm YRZ}$ vs $x$. (b) One quadrant in $k$-space, identifying the critical angles where pseudogap exists. 
The ellipse is a Luttinger pocket from the YRZ theory for $x=0.05$.}
\end{figure}
The electronic specific heat gamma, denoted by $\gamma (T)=C(T)/T$, is
a constant, $\gamma(T)=\frac{2\pi^2}{3} k_{\rm B}^2N(0)$, in the
non-interacting case, with  $N(0)$, the electronic density of states at the Fermi level.
$\gamma(T)$
is presented in Fig.~\ref{fig:fig1}, for several doping values, normalized by the constant $\gamma_{0}=\frac{2\pi^2}{3} k_{\rm B}^2N(0,x=0.16)$.
These results are based on the generic phase diagram, illustrated
by Fig.~\ref{fig:thetas}(a), which is slightly modified from that
previously used by YRZ, with optimal doping now at $x=0.16$. This is more relevant for comparison with the specific heat data of Loram et al.~\cite{loram:1998,loram:1994,loram:2001}  Other phase diagrams based on the analysis of a large database have also appeared in the literature \cite{huefner:2008,tacon:2006}, in which the pseudogap line ends near the upper edge of the superconducting dome, rather than at $x=0.2$.  Since, in this work, we are only interested in making a qualitative comparison with experiment, we have made no attempt to alter other basic parameters of the pseudogap state introduced in the original paper of YRZ\cite{yrz:2006} in order to improve quantitative agreement with experimental data.

Comparison of Fig.~\ref{fig:fig1}(a) with the experimental results of Loram et al.\cite{loram:1998,loram:1994,loram:2001} shows that our theoretical results capture all essential qualitative features observed.  
First, as one proceeds towards the underdoped regime, there is a significant decrease in $\gamma(T)$, as $T_c$ is approached from above.  This reflects an effective decrease in the DOS around the Fermi energy in the normal pseudogapped state.  Second, the jump at $T_c$ is greatly reduced with increased pseudogap.
In Fig.~\ref{fig:fig1}(b), we summarize the doping dependence of the
normalized jump, $\Delta C/\gamma (T_c)T_c$, as well as the normalized
condensation energy, $\Delta U/\Delta U_{\rm n}$ at $T=0$, derived
from our entropy calculations.  Recall that the internal energy, $U$,
is related to the specific heat by $dU/dT=C(T)$ and ultimately to the
entropy.  The condensation energy $\Delta U$ is defined as the
difference between $U$ in the superconducting state, and its value in
the normal state at $T=0$. This is  given by $\Delta
U=\int_{0}^{T_c}(S_{\rm normal}-S_{\rm sc})dT$.  $\Delta U_{\rm n}$ is the condensation energy when the pseudogap is set to zero in both the normal and superconducting states.
\begin{figure}
\includegraphics[width=0.95\linewidth]{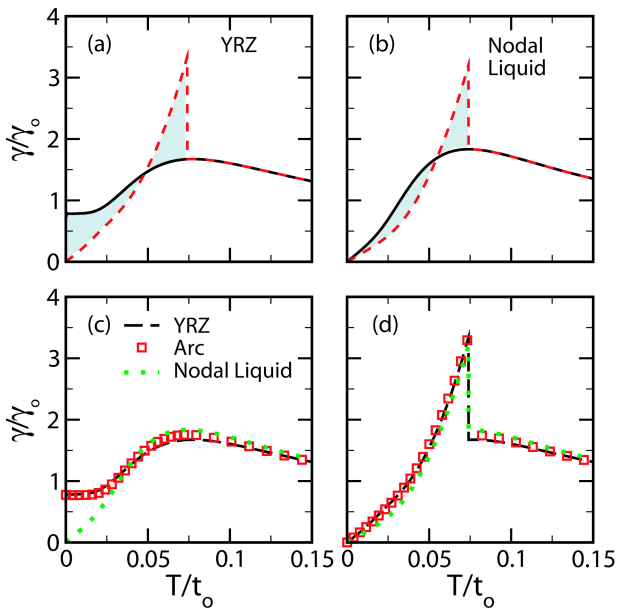}
\caption{\label{fig:areas} (Color online) Plots of $\gamma(T)/\gamma_{0}$
  comparing the superconducting state (red dashed) to the pseudogapped normal
  state (solid), with the areal displacement shaded: (a) YRZ model and (b)
  nodal liquid.  The arc model fitted to YRZ for $x=0.13$ and compared
  to the nodal liquid: (c) pseudogapped normal state and (d) the superconducting state.}
\end{figure}
Both normalized quantities, the jump and the condensation energy, are seen to drop precipitously with decreasing $x$ due to the increase in pseudogap, which is not part of any pure BCS formulation, in which both quantities would be constant for all doping.  Our findings agree qualitatively with the data of Loram et al.\cite{loram:1998,loram:1994,loram:2001}.  It is clear that the model of YRZ has captured an additional essential element of the physics of the underdoped cuprates not present in standard BCS models.  Note that our values of $\Delta C/\gamma(T_c)T_c$ are larger than experiments indicate and that this is mainly due to our use of a large gap ratio of 6, on the order of that indicated by STM.\cite{pasupathy:2008}  However, a smaller value would give better quantitative agreement, indicating that the STM data may not reflect the bulk.

Next, we make connection with the nodal liquid and the Fermi arc approaches.  For both cases, we assume that the pseudogap is located on the usual large
Fermi surface [shown as the solid black curve in Fig.~\ref{fig:thetas}(b)].  This corresponds to replacing the AFBZ energy $\xi_{\boldsymbol{k}}^{0}$ of the YRZ model by the energy $\xi_{\boldsymbol{k}}$.  In this limit, the expression of Eq.~(\ref{eq:entropy}) reduces to the standard BCS expression, with the square of the superconducting gap replaced by the sum of the square of superconducting and pseudogap.  For the Fermi arc model we apply an additional constraint that the pseudogap is non-zero only in an arc located around the antinodal direction, such that
\begin{equation}
\Delta^{\rm ARC}_{\rm pg}(\boldsymbol{k})=\left\{
\begin{array}{ll}\displaystyle
\frac{\Delta_{\rm pg}^{0}}{2}(\cos k_xa -\cos k_ya),  &
(\theta<\theta^{\rm ARC}_{c})\\
\\
\displaystyle
\frac{\Delta_{\rm pg}^{0}}{2}(\cos k_xa -\cos k_ya), &
(\theta>\frac{\pi}{2}-\theta^{\rm ARC}_{c})
\end{array}\right.
\label{eq:pgeqn}
\end{equation}
and $\Delta^{\rm ARC}_{\rm pg}=0$, otherwise. The angle $\theta=\arctan\left[({\pi-k_y})/(\pi-k_x)\right]$ is shown in Fig.~\ref{fig:thetas}(b).
For our purposes, $\theta^{\rm ARC}_{c}$ is a single fitting
parameter, adjusted to give $\gamma_{\rm n}(T=0)$ (n is for the normal pseudogapped state) equal to that of the YRZ model
at the same doping.
 Additionally,
we ignore any temperature dependence of the pseudogap magnitude below
$T_{c}$ as is suggested by Refs.~\cite{storey:CAP:2008} and
\cite{storey:2007} wherein they find that Raman data shows that there
is little modification to the pseudogap below $T_{c}$.  We further ignore complications of the disappearance of the pseudogap at some, relatively high, temperature $T^*$ which is known to cause large specific heat anomalies at $T^{*}$, which are not observed in experiment.
In Fig.~\ref{fig:thetas}(a), the open red squares are the values of
$\theta^{\rm ARC}_c$ obtained from the fit to the specific heat using the construction of Fig.~\ref{fig:thetas}(b). In the YRZ model, the Luttinger pockets also
form an ungapped region analogous to the Fermi arc model and it will
be this region which is responsible for the specific heat. Consequently,
for YRZ, we define $\theta_c^{\rm YRZ}$ as the angle from $(\pi,\pi)$
to the edge of the Luttinger pocket shown in Fig.~\ref{fig:thetas}(b) and plot this in Fig.~\ref{fig:thetas}(a) as a function of doping in comparison with
$\theta^{\rm ARC}$.
  The arc model fits consistently
show $\theta^{\rm ARC}_c<\theta_c^{\rm YRZ}$.  This corresponds to the
additional states (albeit with less quasiparticle weight\cite{yrz:2008,illes:2009}) 
which are located along the AFBZ boundary in the YRZ
model, which are absent in the arc model.  One might imagine
unravelling the Luttinger pocket onto the Fermi surface of the arc
model for a conceptual picture of this comparison. Thus, we expect the
arc model to capture much of the same features as the YRZ model
 for the specific heat.
This contrasts to the nodal liquid case which has the gap over the entire
Fermi surface and no fitting parameters.
    
\begin{figure}
\includegraphics[width=0.95\linewidth]{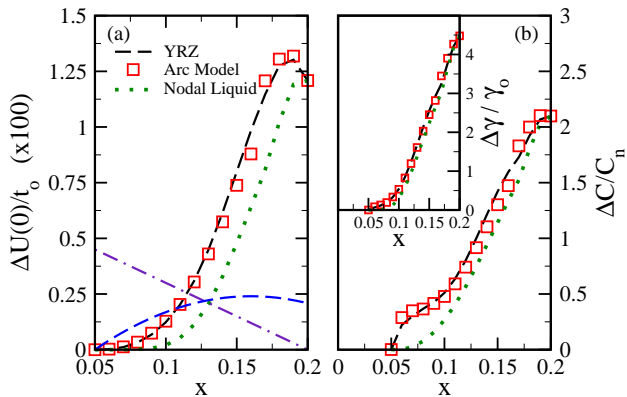}
\caption{\label{fig:fig4}(Color Online) (a) Condensation energy,
  $\Delta U(0)/t_0$, vs $x$ for all three models, overlayed with 
$\Delta_{\rm sc}/t_0$ and $\Delta_{\rm pg}/t_0$ from Fig.~\ref{fig:thetas}. (b) Plot of normalized jump vs $x$ for all three models.  Inset shows $\Delta \gamma (T_c)/\gamma_{0}$.}
\end{figure}

    In the top two frames of Fig.~\ref{fig:areas}, we compare
    superconducting (red dashed curve) and normal pseudogap (solid
    black curve) results for $\gamma(T)$ in the case $x=0.13$.  Frame (a) is for YRZ and frame (b) for the nodal liquid.  The shaded areas illustrate the entropy difference between these two states.  The entropy readjustment is less for the nodal liquid and consequently the specific heat jump at $T=T_c$ is reduced.
These differences arise because the density of states at the Fermi
surface, $N(0)$, is finite in the normal pseudogap state of the YRZ
model while it is zero in the nodal liquid because the pseudogap
exists over the entire Fermi surface, and thus, $\gamma_{\rm
  n}(T\rightarrow0)\rightarrow 0$ (n is for the normal pseudogapped
state).  A further comparison of these two cases is presented in
Fig.~\ref{fig:areas}(c) and \ref{fig:areas}(d), where we have chosen
to compare directly the two normal states and the two superconducting
states, respectively.  The nodal liquid and YRZ agree well at large
$T$, but deviate significantly for $T\lesssim 0.025t_0$, with the
nodal liquid curve going to zero.  The open red squares are results
for the arc model with $\theta^{\rm ARC}_c$ chosen to fit the value of
$\gamma^{\rm YRZ}_{\rm n}(T=0)$.  This results in a finite DOS at the Fermi level, for both the YRZ and arc models, which results in excellent agreement over all temperatures.  Similar results for the superconducting state are presented in Fig.~\ref{fig:areas}(d).  The agreement between YRZ and the arc model is excellent.  Both show some slight deviations from the nodal liquid, but these deviations appear less important than in the normal pseudogapped state results of Fig.~\ref{fig:areas}(c) where we find the low temperature differences to be quite striking.  However, it should be noticed that, in Fig.~\ref{fig:areas}(d), the nodal liquid result (dotted) does fall below YRZ for all $T<T_c$ which results in substantial loss of area under these curves, indicating a smaller condensation energy of the superconducting state [shown in Fig.~\ref{fig:fig4}(a)].  
It is clear from this comparison that placing the pseudogap on the
Fermi surface, rather than on the AFBZ, while at the same time cutting
it off at $\theta^{\rm ARC}_c$ reproduces well the YRZ results. For
YRZ, there is no cut off, but rather the Luttinger contours move away from the gapped AFBZ boundary in the region of the nodes as is clear in Fig.~\ref{fig:thetas}(b).  

Fig.~\ref{fig:fig4} provides further comparison of both the nodal liquid and Fermi arc models with the YRZ results, across the entire doping range.  Fig.~\ref{fig:fig4}(a) compares the condensation energy $\Delta U$ vs doping, $x$.  There is excellent agreement between YRZ and the arc model, while the nodal liquid is consistently lower.  Although the nodal liquid still captures the large decrease in condensation energy that is caused by the opening of the pseudogap, it overestimates the effect.
The same remarks apply to Fig.~\ref{fig:fig4}(b) where $\Delta C/C_n$ is shown vs $x$.  In the inset, we show the $\Delta \gamma (T_c)/\gamma_{0}$ for completeness.

In summary, we have found that the microscopic model of YRZ, based on the RVB spin liquid, which includes as its central essential element the formation of a pseudogap on the AFBZ in the underdoped region of the cuprate phase diagram, can account for all of the qualitative characteristics of the observed evolution of the specific heat as a function of doping.  We have also found that an arc model with pseudogap formation on the Fermi surface itself, but limited to a region around the antinodal direction resulting in an ungapped arc beyond $\theta_c$ in the nodal region, can adequately simulate the results obtained in YRZ theory.  The basic reason for this fortunate circumstance is that, in YRZ, the Luttinger surfaces define pockets in the nodal direction around, but not directly on the antiferromagnetic Brillouin zone boundary, which keeps the electrons in that nodal region from fully sampling the pseudogap on the AFBZ.  We have shown that this effect can be well approximated by the ungapped Fermi arcs of the arc model.  
On the other hand, the nodal liquid idea with pseudogap over the entire Fermi surface accentuates the pseudogap effect as compared to YRZ.  Nevertheless, such a model still has merit; because of its great simplicity, it can give straightforward insight into the qualitative behaviour of the pseudogapped state.

\begin{acknowledgments}
  This work has been supported by the Natural Sciences and
Engineering Council of Canada (NSERC) and the Canadian Institute
for Advanced Research (CIFAR).
\end{acknowledgments}

\bibliographystyle{apsrev}
\bibliography{bib}

\end{document}